\documentclass[11pt,twoside]{article}
\usepackage{asp2004}
\usepackage{psfig}
\usepackage{epsf}
\usepackage{graphics}
\usepackage{lscape}
\def\edcomment#1{\iffalse\marginpar{\raggedright\sl#1\/}\else\relax\fi}
\reversemarginpar

\begin{document}
\ifx\href\undefined\else\hypersetup{linktocpage=true}\fi
\title{Extragalactic Ultracompact HII Regions:\\ 
Probing the Birth Environments of Super Star Clusters} 
\author{Kelsey E. Johnson} 
\affil{Astronomy Dept.,
University of Wisconsin, 475 N. Charter St., Madison, WI, U.S.A.}

\begin{abstract}
In recent years, a number of extragalactic massive star clusters that
are still deeply embedded in their birth material have been
discovered.  These objects represent the youngest stage of massive
star cluster evolution yet observed, and the most massive and dense of
these may be proto globular clusters.  Their properties appear to be
similar to those of ultracompact HII regions in the Galaxy, but scaled
up in total mass and luminosity.  In many cases, these clusters are
only visible at mid-IR to radio wavelengths, and they have typically
been detected as ``inverted'' spectrum radio sources.  However, the
set of existing observations is anemic, and our current physical model
for these natal clusters in simplistic.  This article will overview
what we think we know about these objects based on existing
observations and outline some of the most significant gaps in our
current understanding.
\end{abstract}
\thispagestyle{plain}

\section{Introduction}

\begin{figure}
\plotfiddle{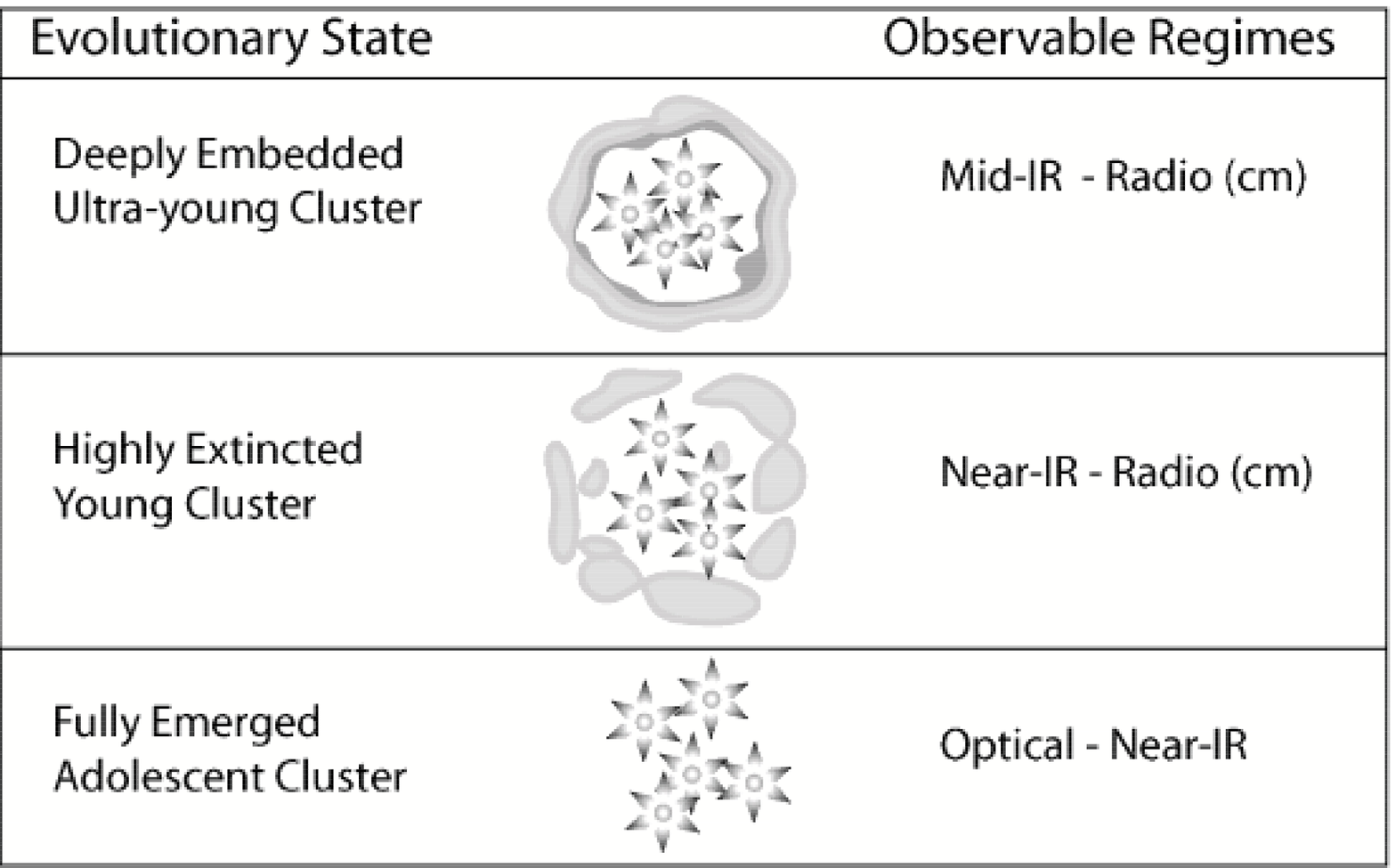}{7cm}{0}{60}{60}{-160}{-0}
\caption{A cartoon illustrating stages in the early evolution of massive
star clusters.  Observations in at thermal infrared to radio 
wavelengths are required to probe their birth environments. \label{evolv} }
\end{figure}

There is no question that massive star clusters, and super star
clusters in particular, are among the most extreme star formation
environments in the local universe.  Moreover, these clusters can have
a dramatic affect on their surrounding interstellar medium, and in
some cases the intergalactic medium.  Nevertheless, the conditions
required for their formation remain far from understood.  As just one
example, there is still significant debate over whether super star
clusters (SSCs) are merely the result of populating a cluster initial
mass function statistically, or whether they truly require a separate
mode of star formation.  While a multitude of adolescent massive star
clusters have been observed at optical wavelengths, these observations
cannot penetrate the natal material surrounding the youngest clusters
in order to reveal the properties of their birth environments
(Figure~\ref{evolv}).  Nevertheless, if we want to understand the
formation of massive star clusters, it is not a terribly bad idea to
observe them while they are forming.

A number of investigations at infrared (IR) and radio wavelengths have
now revealed a population of extremely young massive star clusters in
a growing sample of galaxies \citep[e.g.][]{turner98, kj99, tarchi00,
beck01, gorjian01, vacca02, johnson01, neff00, plante02, beck02,
johnson03, jk03}.  These observations suggest that the early stages of
massive star cluster evolution may parallel those of individual
massive stars observed in the Milky Way.  Therefore, our knowledge
about individual ultracompact H{\sc ii} regions (UCH{\sc ii}s) is
critical to understanding and interpreting observations of embedded
massive star clusters.  Of course, individual massive star formation
is also poorly understood, and we would be remiss to blindly
extrapolate the inferred properties of individual massive proto-stars
to the formation of SSCs.  Perhaps the most obvious way in which this
extrapolation could lead us astray is by neglecting the effects of the
density of proto-stars on star formation.  In fact, it seems
likely that a tremendous amount of physical insight will come from
trying to understand how the properties of star formation scale
between the two extremes of {\it individual} massive star formation
and {\it SSC} formation.

Despite the pioneering observations done by a number of investigators,
our current knowledge about embedded massive star clusters is anemic,
and we have only begun to investigate their properties, environments,
and evolution.  In the pages that follow, I will discuss the current
observational limitations (\S\ref{obs_limits}), give a brief tour of
the sample of extragalactic embedded clusters (\S\ref{sample}),
overview what we think we know about these objects (as well as what we
do not know, \S \ref{knowledge}), and outline the impact that the next
generation of powerful observatories at infrared to radio wavelengths
will have on this area of research (\S\ref{future}).

\section{Observational Limits \label{obs_limits}}

\begin{figure}[!ht]
\plotfiddle{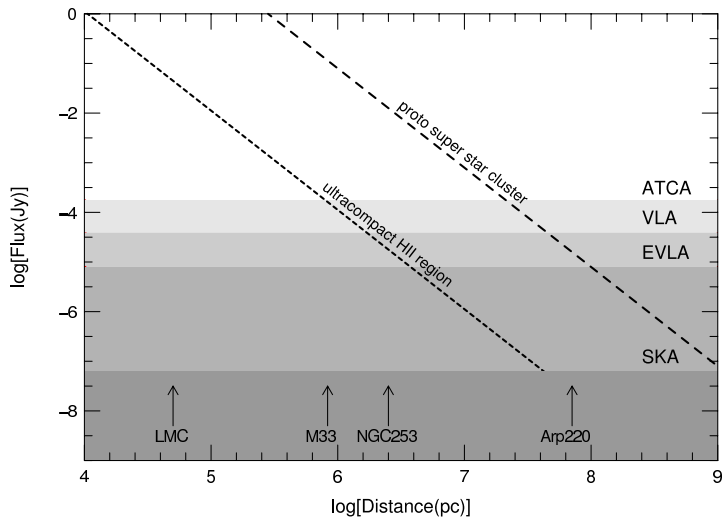}{7cm}{0}{39}{39}{-160}{-15}
\caption{The 5$\sigma$ sensitivities for 8~hr integrations at 6~cm for
various radio interferometers (both current and planned) are indicated
by the horizontal grey shaded areas. The flux densities of ``typical''
UCH{\sc II}s and natal super star clusters as a function of
distance are shown with the diagonal lines. The distances to a handful
of reference galaxies are indicated. \label{radio_obs}}
\end{figure}

Because star formation takes place within dense and dusty cocoons, we
can only probe these regions with observations at infrared or longer
wavelengths.  In fact, some natal clusters (presumably the youngest)
are even invisible at near-IR wavelengths as long as the K-band
\citep[$\sim 2.2 \mu$m,][]{vacca02} or even the L-band ($\sim 3.8
\mu$m, M. Sauvage, private communication).  On the other hand, radio
emission is completely unfettered by dust, but the observations need
to be sensitive at high frequencies ($> 5$~GHz) in order to avoid
self-absorption by the dense gas and disentangle non-thermal radio
emission.  Of course, obtaining the necessary sensitivity is only half
of the battle; achieving adequate spatial resolution is also a
limiting constraint with current facilities.  Roughly speaking, a
linear resolution of at least $\sim 0.5$~pc is necessary to observe
individual UCH{\sc ii}s, and embedded SSCs require a linear resolution
of $\sim 10$~pc or better.

In principle, mid-IR to sub-millimeter wavelengths are ideal for
observing extragalactic UCH{\sc ii}s and natal SSCs because these
objects are extremely luminous in this regime due to thermal dust
emission (Figure~\ref{future_inst}).  As a demonstration of their
potency at thermal wavelengths, UCH{\sc ii}s are among the most
luminous Galactic objects in the {\it IRAS} catalog \citep[$>60$\% of
the {\it IRAS} sources with flux densities $>10^4$~Jy at 100$\mu$m are
UCH{\sc ii}s,][]{wc89}, and the embedded SSCs in the starburst
galaxies He~2-10 and NGC~5253 are responsible for nearly all of the
mid-IR emission from these galaxies \citep{beck01, gorjian01,
vacca02}.  However, we are severely limited by spatial resolution with
the facilities currently available at infrared to
sub-millimeter wavelengths.  This situation will improve to some
degree with the next generation of observatories that are
beginning to come on-line; however, even the {\it Spitzer, Herschel,}
and {\it SOFIA} telescopes will not obtain spatial resolutions of better
than $\sim 1''$ at their shortest wavelengths.

The sensitivity and spatial resolution of existing radio
interferometers is somewhat better matched to observing extragalactic
UCH{\sc ii}s and natal SSCs.  Figure~\ref{radio_obs} illustrates the
sensitivities of various radio interferometers.  The current
facilities have the sensitivity to detect individual UCH{\sc ii}s
nearly as far away as the starburst galaxy NGC~253, or embedded SSCs
nearly out to ultraluminous infrared galaxy Arp~220.  Perhaps the most
obvious places to look for extragalactic UCH{\sc ii}s are the
Magellanic Clouds, however the best spatial resolution available in
the southern hemisphere from the {\it Australia Telescope Compact
Array} at this distance is $\sim 1$~pc, which limits our ability to
asses whether H{\sc ii} regions are actually ``ultra'' compact.
Another obvious place to search of UCH{\sc ii}s is in M33, the sister
galaxy to the Milky Way. However, a linear resolution of better than
$\sim 0.1''$ is required in order to observe UCH{\sc ii}s at this
distance (840~kpc), which will be difficult (but not impossible) to
achieve with the {\it Very Large Array} (VLA).  The constraints are
somewhat more relaxed for observing embedded SSCs, which are both
significantly larger and brighter than individual UCH{\sc ii}s.  With
the {\it VLA}, it is possible to resolve embedded SSCs in galaxies
out to distances of $\sim 10$~Mpc.  Beyond this distance, both
crowding and contamination by non-thermal and optically-thin thermal
radio emission becomes an issue, and one has to use caution when
interpreting the data.

\begin{figure}
\plotfiddle{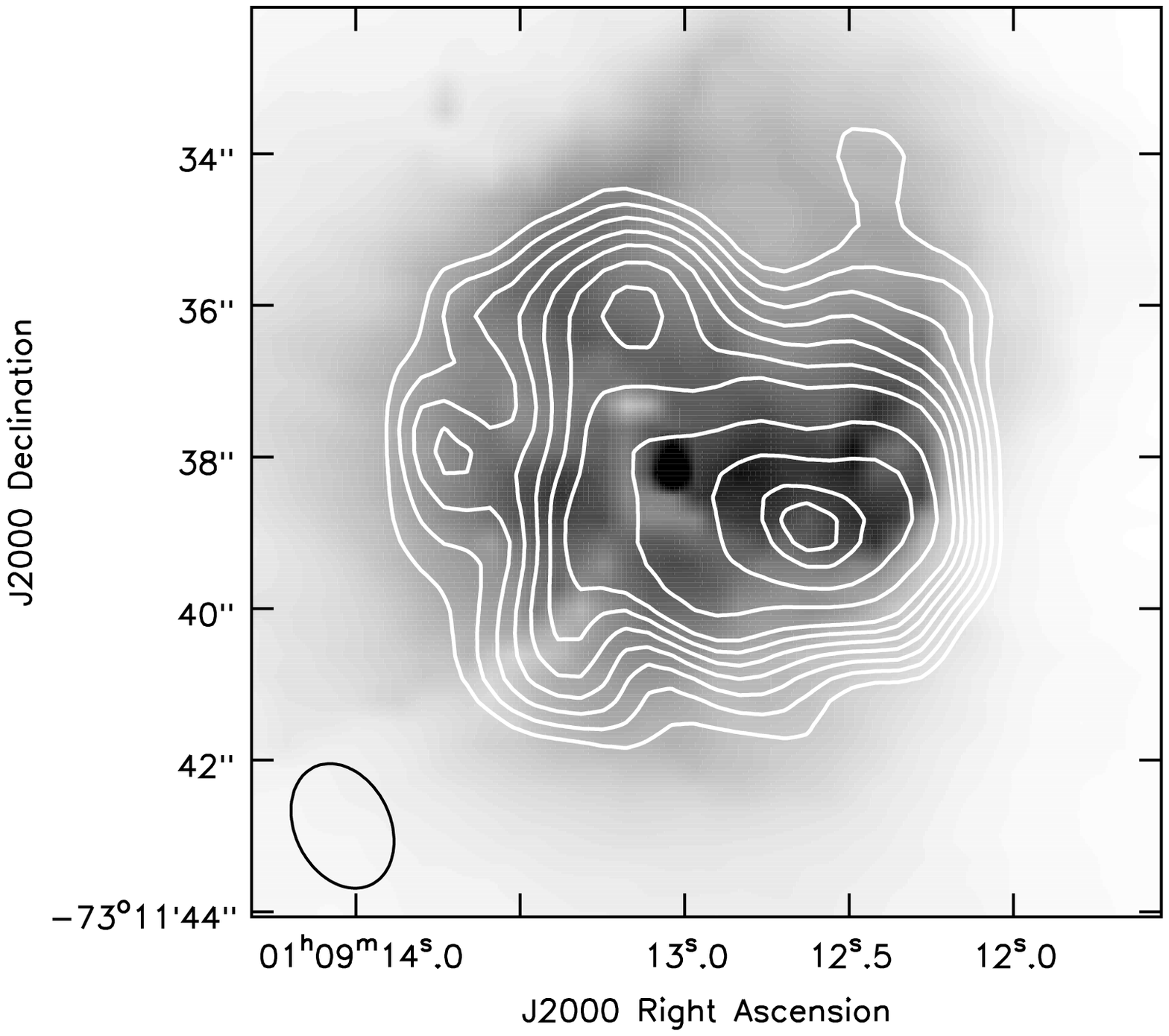}{4.2cm}{0}{38}{38}{-205}{-100}
\plotfiddle{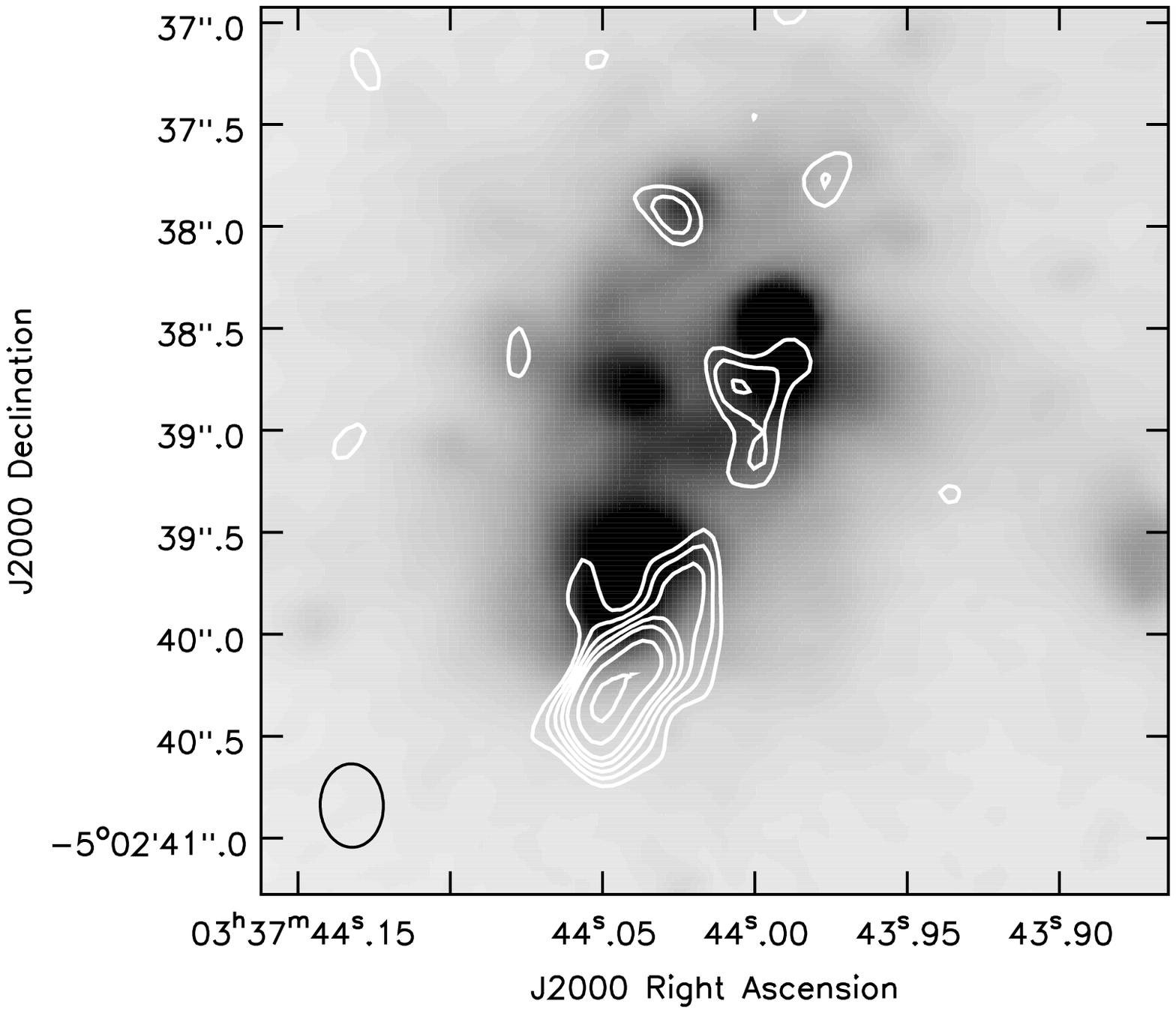}{0cm}{0}{38}{38}{-20}{-75}

\caption{Examples of the most nearby and most distant embedded
clusters yet studied. {\bf (left)} The H{\sc ii} region N81 in the SMC;
an {\it HST} H$\alpha$ image is shown in gray-scale, and {\it ATCA} 3~cm radio
emission is shown in contour \citep{indebetouw04}.  {\bf (right)} The
starburst galaxy SBS0335-052; an {\it HST} I-band image is shown in
gray-scale, and {\it VLA} 1.3~cm radio emission is shown in contour (Johnson
\& Plante in prep.)
\label{examples} }
\end{figure}

\section{Sample of Extragalactic Natal Clusters \label{sample}}
A number of embedded massive star clusters have been identified in a
variety of galaxies.  The objects that have been most well-studied to
date are those that were identified in the incipient radio
observations of the starburst galaxies NGC~5253 ($\sim 4$~Mpc) and
He~2-10 ($\sim 9$~Mpc) \citep{turner98, kj99}.  Subsequent work has
revealed a number of embedded star forming regions in a range of
galactic environments.  In the most nearby examples,
\citet{indebetouw04} have identified a number of candidate UCH{\sc
ii}s and UCH{\sc ii} complexes in the Magellanic Clouds (here the word
``candidate'' is used because the spatial resolution is not
adequate to determine whether the regions are ``ultra'' compact).  In
this sample, the radio observations indicate stellar ionizing fluxes
and morphologies consistent with sources ranging from single UCH{\sc
ii}s to relatively modest UCH{\sc ii} associations
(Figure~\ref{examples}a).

Compact radio-detected clusters have also been identified in galaxies
such as NGC~6822 at $\sim 0.5$~Mpc, IC~4662 at $\sim 2$~Mpc, IIZw40 at
$\sim 11$~Mpc, Haro~3 at $\sim 13$~Mpc, NGC~5398 at $\sim 17$~Mpc,
NGC~2146 at $\sim 15$~Mpc, the Antennae at $\sim 21$~Mpc, and
SBS~0335-052 at $\sim 53$~Mpc 
\citep[][Johnson \& Plante in prep.,
Seth \& Johnson in prep.]{neff00, tarchi00, beck02, johnson03, johnson04}
\footnote{This list in not intended to be exclusive or exhaustive.}.
The star forming regions in this sample have properties ranging from
those of natal OB-associations to incredibly massive SSCs.  For
example, the main radio source in SBS~0335-052 has an ionizing flux
equivalent to $\sim 12,000$~O7-type stars (Figure~\ref{examples}b;
note that the linear resolution of these observations is only $\sim
75$~pc, and this source is likely to be composed of several individual
clusters).

\section{What We Think We Know \& What We Know We Don't Know \label{knowledge}}

Given the limited observations available at the present time, attempts
to constrain the physical properties of embedded clusters have been
fairly crude.  As a result, the current physical model we have for
these objects is simplistic: When an SSC is born, the
massive stars ionize the surrounding natal material,
creating an extremely dense H{\sc ii} region.  The dense H{\sc ii}
region is in turn surrounded by a dust cocoon that is presumably quite
warm on the inner boundary, with the temperature decreasing at
distances farther from the ionizing stars.  The constituent ionizing
stars may be surrounded by individual H{\sc ii} regions and dust
cocoons in the earliest stages, but this is currently unknown.
Clusters that are more sparse are likely to resemble ultracompact
H{\sc ii} region (UCH{\sc ii}) complexes in the Milky Way (such as
W49A), and contain spatially discrete H{\sc ii} regions and cocoons,
but the existence of individual cocoons becomes less tenable in the
extreme stellar densities of the cores of massive SSCs.  

The dense H{\sc ii} regions are characterized by an ``inverted'' radio
spectral energy distributions due to optically thick free-free
emission ($\alpha >0$, where $S_\nu \propto \nu^\alpha$).  The
frequency at which the radio emission becomes optically thick is
largely dependent on the density of the H{\sc ii} region.  In the
earliest stages of cluster evolution, all of the stellar luminosity is
reprocessed by the surrounding dust cocoon, and this cocoon is
observable at thermal infrared to sub-millimeter wavelengths.  To some
extent, the nature of the resulting spectral energy distribution
depends on whether the dust in the cocoon is smooth or clumpy, but
this is not currently constrained by observations.

\begin{figure}
\plotfiddle{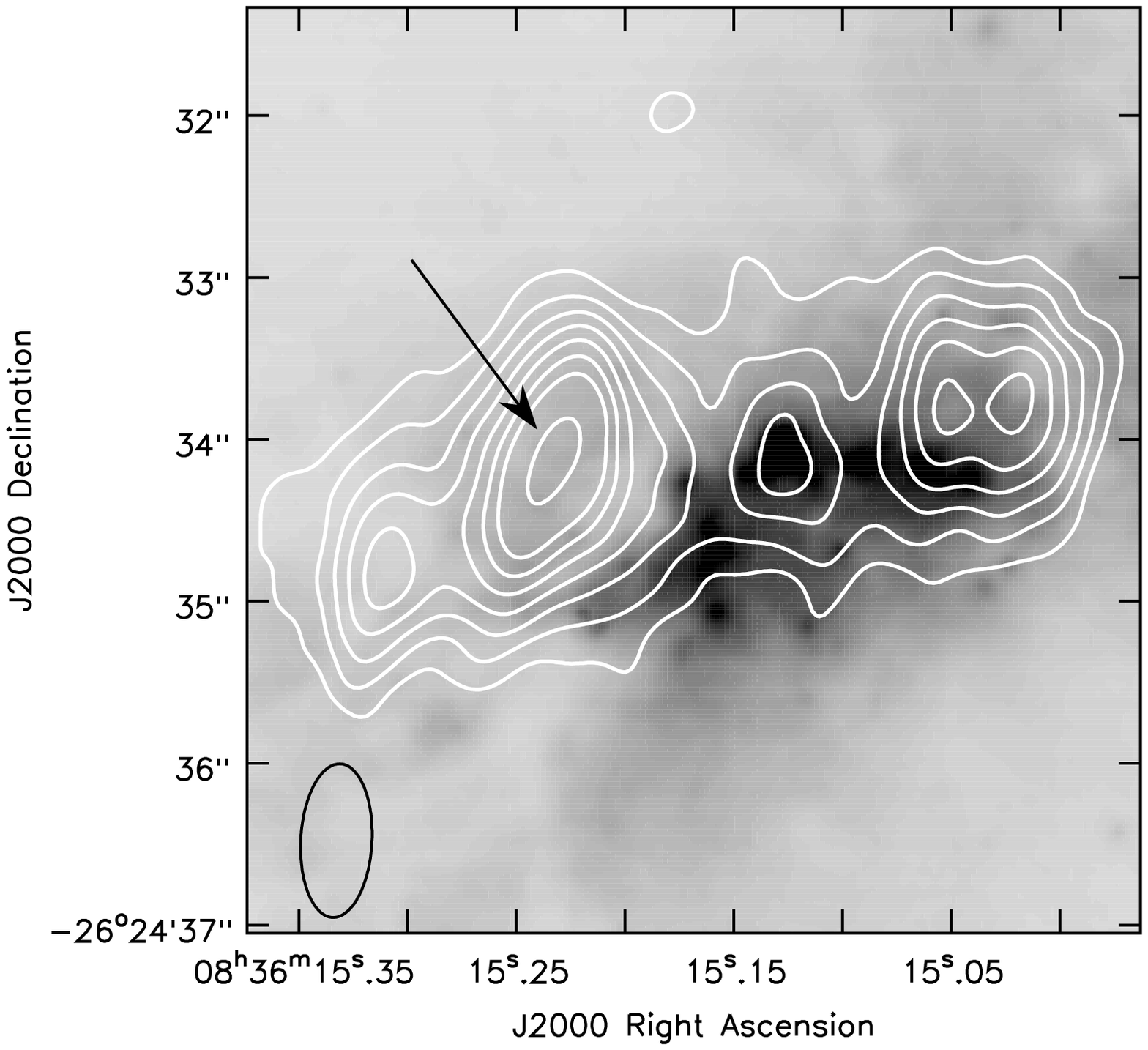}{3.8cm}{0}{32}{32}{-212}{-75}
\plotfiddle{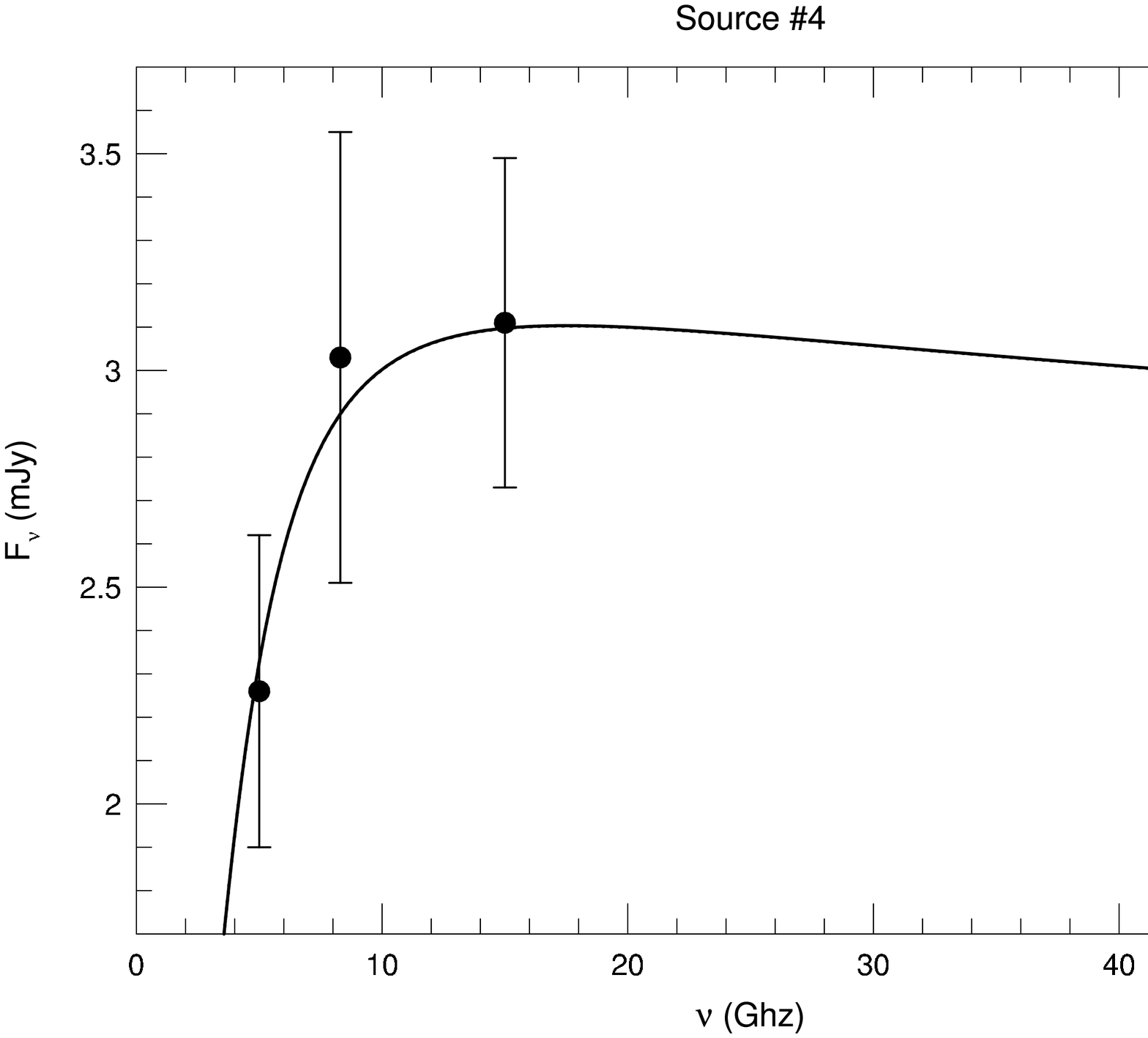}{0cm}{0}{28}{28}{-5}{-10}
\caption{
{\bf (left)} Natal SSCs in the starburst galaxy He~2-10.
An {\it HST} I-band image is shown in gray-scale, and {\it VLA} 2cm emission
is shown in contour.  The four radio sources without optical counterparts are
natal SSCs. The arrow indicates the source for which the spectral energy 
distribution is shown.
{\bf (right)} The radio fluxes for the most luminous
natal SSC in He~2-10 along with the best fit spectral energy distribution 
for a model H{\sc ii} region.  
\label{He2_10} }
\end{figure}

Existing observations have provided {\it estimates} for some of the
physical properties of natal clusters, including size, electron
density, pressure, stellar mass, H{\sc ii} mass, dust mass,
and age.  For the most nearby clusters, their size can be directly
measured using high spatial resolution observations.  For more distant
clusters, model H{\sc ii} regions can be used to fit the radio
spectral energy distribution and infer both the radius and electron
density (Figure~\ref{He2_10}b).  The embedded SSCs that have been
analyzed in this way \citep[e.g. those in He~2-10,][]{jk03} have H{\sc
ii} region radii of only a few parsecs, and the electron densities
have global values of $\sim 10^3$~cm$^{-3}$, with peak values in
excess of $\sim 10^6$~cm$^{-3}$.  Assuming temperatures of $\sim
10^4$~K, these densities imply global pressures of $P/k_B > 10^7$~K~cm$^{-3}$
and peak pressures reaching values of $P/k_B \sim 10^{10}$~K~cm$^{-3}$.

The masses associated with the stellar clusters, H{\sc ii} regions, and
dust cocoons can be bootstrapped from radio and infrared observations.
The optically-thin thermal flux (typically measured with the highest
available radio frequency) can be used to infer a total ionizing
luminosity, and the ionizing luminosity can be translated into an
embedded stellar mass by assuming a stellar initial mass function.
Using this technique, clusters with a range of masses have been
detected (spanning modest OB-associations to massive SSCs).  The H{\sc
ii} mass can be estimated given the radii and densities for the
ionized regions.  In at least some cases, the H{\sc ii} mass has
anomalously low values \citep[$< 5$\% of the stellar mass,][]{jk03},
which we tentatively interpret as a sign of the youth of the H{\sc
ii} regions.  The mass of the dust cocoons around embedded SSCs has
also been estimated in a few cases using infrared observations; these
estimates have suggested dust masses of $\sim 10^5 - 10^6
M_\odot$ for massive SSCs in SBS0335-052 and He~2-10 \citep{plante02, vacca02}.

The ages of the natal clusters are exceptionally tricky to ascertain,
as there are no available spectral diagnostics to determine age at
wavelengths longer than the near-IR (and for shorter wavelengths the
diagnostics are unreliable for ages less than $\sim 1$~Myr in any
case).  The best we can do is offer circumstantial evidence and look
at the back of a proverbial envelope.  There are two independent
``back-of-the-envelope'' calculations that can be used to constrain
the lifetimes of the embedded phase of SSC evolution: (1) The
pressures inferred for these regions are several orders of magnitude
higher than those typical in the ISM.  As a result, the H{\sc ii}
regions ought to expand and dissipate on very short time-scales ($\tau
< 10^6$~yrs).  However, there are caveats to this argument: First, we
do not know the pressure of the surrounding ISM, and it could be a
good deal higher than normal since it is a starburst region.  Second,
if the clusters are massive enough, the gravity could be large enough
to help balance pressure and confine the H{\sc ii} region
\citep[e.g.][]{turner03}. (2) If star formation has been relatively
continuous over $\sim 10$~Myr, then a comparison between the total
number of clusters with ages $< 10$~Myr and the number embedded
clusters ought to reflect the their relative lifetimes.  When this has
been applied to galaxies with good enough statistics
\citep[e.g. He~2-10,][]{kj99}, the clusters appear to spend $\sim
10-20$\% of the lifetimes of the constituent massive stars in the
embedded phase ($\tau \sim 0.5 -1$~Myr).  There are caveats to this
argument as well.  First, it relies on star formation being continuous
over the lifetimes of massive stars.  Second, we have to make sure
that both the optical and radio observations are sensitive to roughly
the same cluster masses.  In any case, it is comforting that both
calculations suggest the same lifetimes for the embedded phase of
cluster evolution, so our estimates are probably not too far off.

While we have estimates for a few of the quantities associated with
the birth of massive star clusters, there are a potpourri of
properties that it would be nice to know more about.  For example, it
would be instructive to {\it directly} measure densities, pressures,
and temperatures using various line diagnostics at IR to radio
wavelengths.  Another quantity that might be particularly useful for
theorists (that is almost completely unconstrained) is the star
formation efficiency in these objects.  We also have no idea how much
ionizing radiation could be escaping from the natal cocoons (and it
could be quite a lot if the dust is clumpy).  In fact, we know almost
nothing about the natal dust cocoons (including the temperatures,
grain compositions, and geometry).  This list is certainly not
exhaustive.  An understanding of these quantities, and how they relate
to the properties of the resulting cluster, will eventually provide a
great deal of insight into the requirements for massive star cluster
formation.

\section{Future Prospects \label{future}}

\begin{figure}
\plottwo{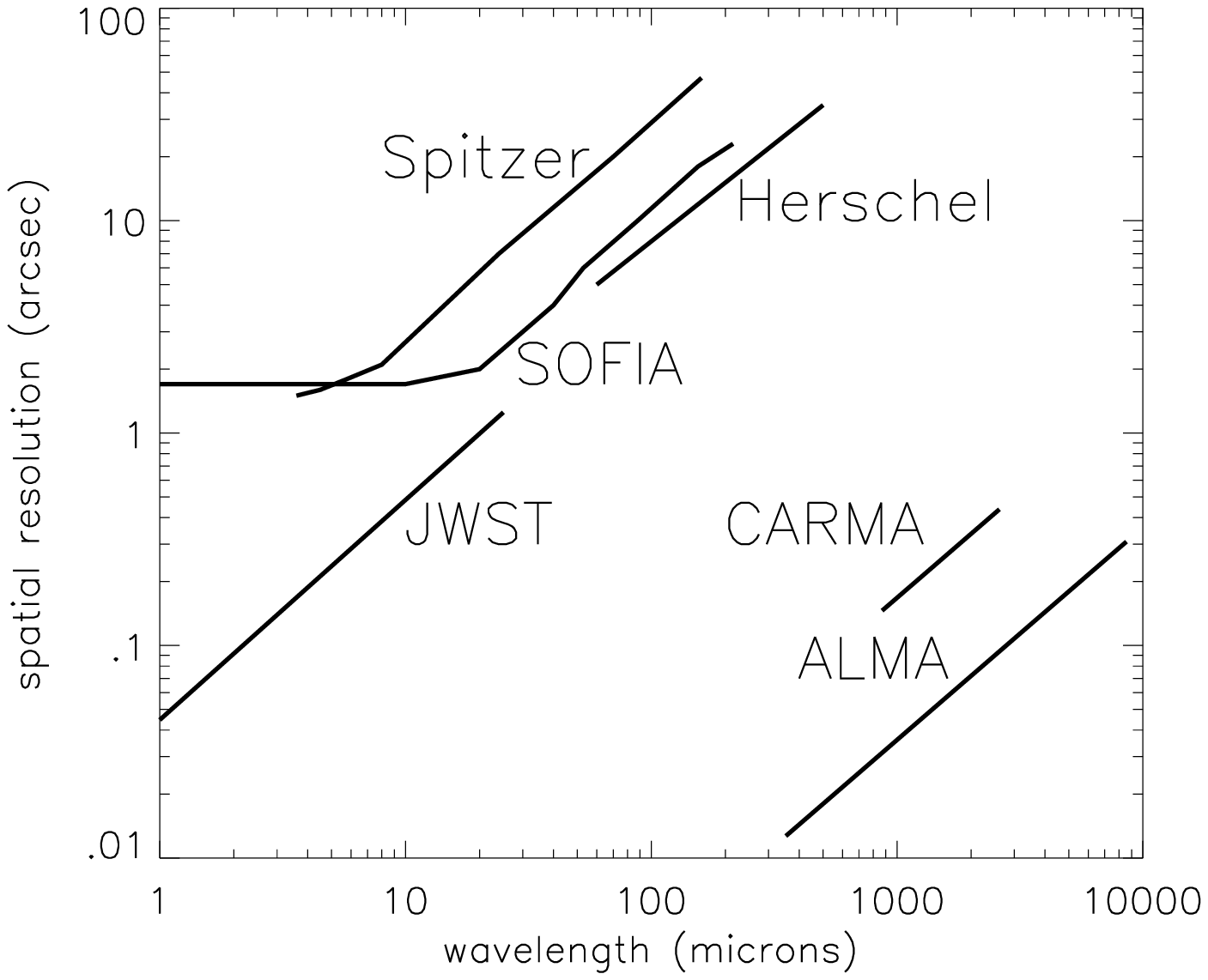}{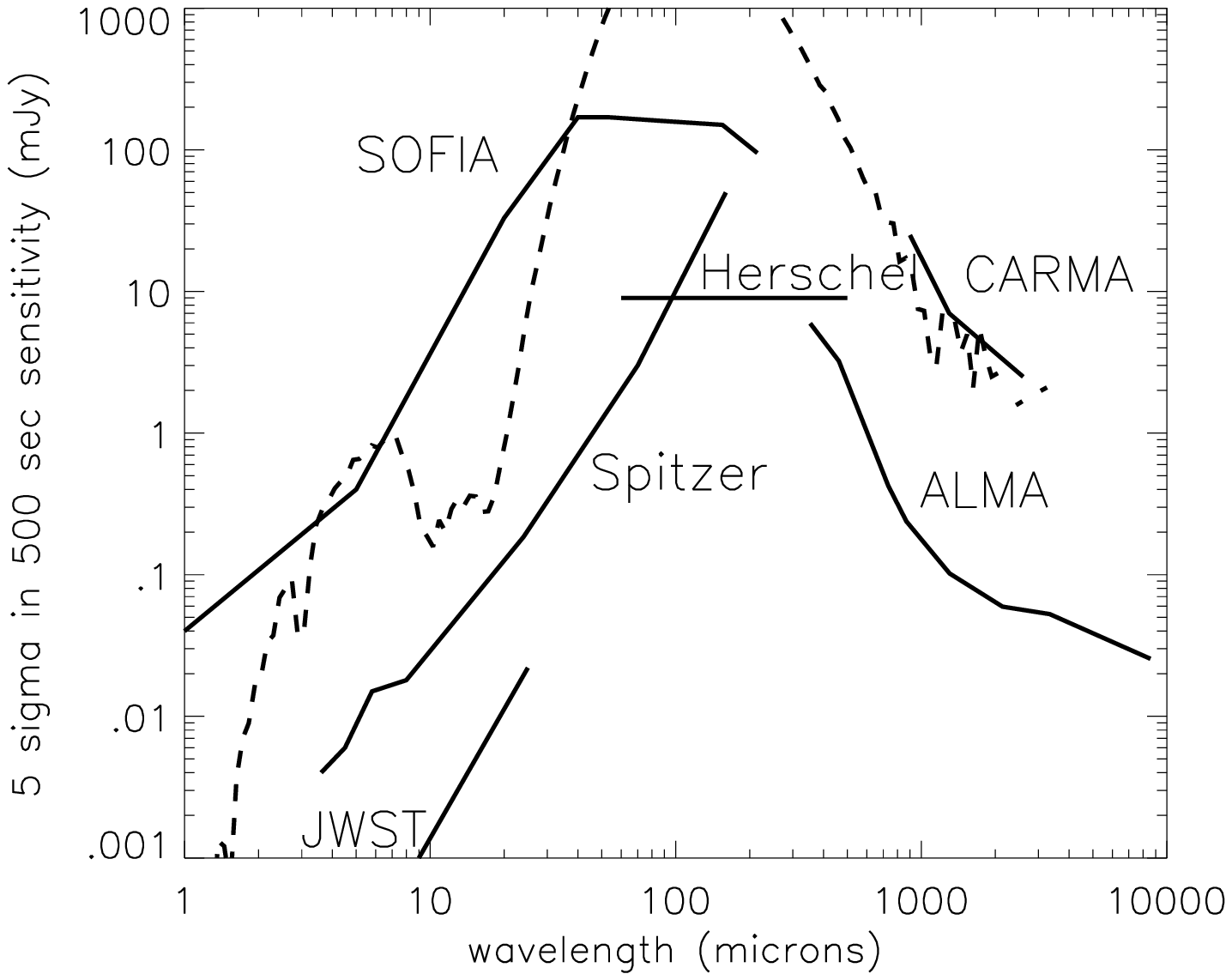}
\caption{{\bf (left)} The anticipated spatial resolution and
wavelength coverage for upcoming infrared and millimeter
observatories. {\bf (right)} The sensitivities of these observatories
normalized to 500s integration time.  A model spectral energy
distribution (Whitney et al. in prep) for a $10^6 M_\odot$ natal SSC
at a distance of 10~Mpc is over-plotted with a dashed line for
reference. \label{future_inst} }
\end{figure}

While there are no lack of questions remaining about the birth of massive
star clusters, it is an exceptional good time to be asking these
questions.  Over the next decade (or so), a range of powerful
facilities are going to become available at infrared to radio
wavelengths.  Figure~\ref{future_inst} illustrates the spatial
resolution and sensitivity of the IR to millimeter
facilities that will soon become available.  In the thermal infrared,
{\it Spitzer, Herschel}, and {\it SOFIA} will provide a vast
improvement over previous facilities, and embedded star forming
regions are likely to be a dominate source of flux in the thermal
spectral energy distribution of star forming galaxies. However, we
will have to be cautious in our interpretation of data from these
observatories because they will not have the spatial resolution
necessary to isolate individual natal super star clusters.  At
millimeter wavelengths, the {\it Atacama Large Millimeter Array} will
open up an entirely new window for observing extragalactic star
formation with resolution and sensitivity incredibly well-matched to
observing embedded clusters.  In the radio regime (see
Figure~\ref{radio_obs}), the {\it Expanded Very Large Array} and {\it
Square Kilometer Array} are going to achieve sensitivities and
resolutions that will allow us to study nearby natal clusters in
fantastic detail and enable us to discover new natal clusters out to
distances of $100$~Mpc and beyond.

\acknowledgments{This work has benefited from many fruitful discussions
with my collaborators, including P.Conti, M.Goss, R.Indebetouw, H.Kobulnicky,
and W.Vacca.
I also gratefully acknowledge support for this work
provided by the NSF through an Astronomy and Astrophysics Postdoctoral
Fellowship.  Support for proposal \#09934 was provided by NASA through
a grant from the Space Telescope Science Institute, which is operated
by the Association of Universities for Research in Astronomy, Inc.,
under NASA contract NAS5-26555.  }


\begin{thebibliography}{}

\bibitem[Beck et al.(2002)]{beck02}
Beck, S.C., Turner, J.L., Langland-Shula, L.E., Meier, D.S., Crosthwaite, L.P.,
Gorjian, V. 2002, \aj\, 124, 2516

\bibitem[Beck, Turner, \& Gorjian(2001)]{beck01}
Beck, S.C., Turner, J.L., \& Gorjian, V. 2001, \aj\, 122, 1365

\bibitem[Gorjian, Turner, \& Beck(2001)]{gorjian01}
Gorjian, V., Turner, J.L., \& Beck, S.C. 2001, \apj\, 554

\bibitem[Indebetouw, Johnson, \& Conti(2004)]{indebetouw04}
Indebetouw, R., Johnson, K.E., \& Conti, P.S. 2004, \aj\, submitted

\bibitem[Johnson et al.(2004)]{johnson04} Johnson, K.E., Indebetouw, R., 
Watson, C., \& Kobulnicky, H.A. 2004, AJ, submitted

\bibitem[Johnson \& Kobulnicky(2003)]{jk03}
Johnson, K.E. \& Kobulnicky, H.A. 2003, \apj\, 597, 923

\bibitem[Johnson, Indebetouw, \& Pisano(2003)]{johnson03}
Johnson, K.E., Indebetouw, R., \& Pisano, D.J. 2003, \aj\, 126, 101

\bibitem[Johnson et al.(2001)]{johnson01}
Johnson, K.E., Kobulnicky, H.A., Massey, P., \& Conti, P.S. 2001, \apj\, 559,
864

\bibitem[Kobulnicky \& Johnson(1999)]{kj99}
Kobulnicky, H.A. \& Johnson, K.E. 1999, \apj\, 527, 154

\bibitem[Plante \& Sauvage(2002)]{plante02}
Plante, S. \& Sauvage, M. 2002, \aj\, 124, 1995

\bibitem[Neff \& Ulvestad(2000)]{neff00}
Neff, S.G. \& Ulvestad, J.S. 2000, \aj\, 120, 670

\bibitem[Tarchi et al.(2000)]{tarchi00}
Tarchi, A., Neininger, N., Greve, A., Klein, U., Garrington, S.T., Muxlow, 
T.W.B., Pedlar, A., Glendenning, B.E. 2000, \aap\, 358, 95

\bibitem[Turner et al.(2003)]{turner03}
Turner, J.L., Beck, S.C., Crosthwaite, L.P., Larkin, J.E., McLean, I.S., Meier, 
D.S. 2003, Nature, 423, 621

\bibitem[Turner, Ho, \& Beck(1998)]{turner98}
Turner, J.L., Ho, P.T.P., \& Beck, S.C. 1998, \aj\, 116, 1212

\bibitem[Vacca, Johnson, \& Conti(2002)]{vacca02}
Vacca, W.D., Johnson, K.E., \& Conti, P.S. 2002, \aj\, 123, 772

\bibitem[Wood \& Churchwell(1989)]{wc89}
Wood, D.O. \& Churchwell, E. 1989, \apjs\, 69, 831

\end{thebibliography}
\end{document}